\documentclass[12pt]{iopart}

\expandafter\let\csname equation*\endcsname\relax
\expandafter\let\csname endequation*\endcsname\relax
\DeclareUnicodeCharacter{0301}{\'{e}}

\usepackage[utf8]{inputenc}
\usepackage[american,british]{babel}
\usepackage[T1]{fontenc}
\usepackage[pdftex]{graphicx}  
\usepackage{graphicx, xcolor}
\usepackage{dcolumn}
\usepackage{braket}
\usepackage{bm}
\usepackage{amsmath,amsthm,amssymb}
\usepackage{color}
\usepackage{verbatim}
\usepackage{lipsum}
\usepackage[colorlinks,linkcolor=blue,citecolor=blue,urlcolor=blue]{hyperref}

\usepackage{physics}
\usepackage{mathtools}

\usepackage[backend=bibtex, style=numeric-comp, doi=true, isbn=false, url=false, eprint=false, sorting=none,maxnames=10,firstinits=true]{biblatex}
\begin{document}

\title[Partial and infinite-temperature thermalization  on a quantum hardware]{Observation of partial and infinite-temperature thermalization induced by repeated measurements on a quantum hardware}

\author{Alessandro Santini$^{1}$, Andrea Solfanelli$^{1,2}$, Stefano Gherardini$^{1,4,5}$, Guido Giachetti$^{6}$}

\address{$^1$SISSA, via Bonomea 265, 34136 Trieste, Italy}
\address{$^2$INFN Sezione di Trieste, via Bonomea 265, 34136 Trieste, Italy}
\address{$^3$Center for Life Nano-Neuroscience, Italian Institute of Technology,
viale Regina Elena 291, 00161, Roma, Italy}
\address{$^4$CNR, Istituto Nazionale di Ottica, Area Science Park, 34149 Trieste, Italy}
\address{$^5$The Abdus Salam International Center for Theoretical Physics (ICTP), Strada Costiera 11, I-34151 Trieste, Italy}
\address{$^6$LPTM, Universit\'e Paris Cergy, Av. Adolphe Chauvin 2, 95300 Pontoise, France}

\ead{asantini@sissa.it}

\date{\today}

\begin{abstract}
    On a quantum superconducting processor we observe partial and infinite-temperature thermalization induced by a sequence of repeated quantum projective measurements, interspersed by a unitary (Hamiltonian) evolution. Specifically, on a qubit and two-qubit systems, we test the state convergence of a monitored quantum system in the limit of a large number of quantum measurements, depending on the non-commutativity of the Hamiltonian and the measurement observable. When the Hamiltonian and observable do not commute, the convergence is uniform towards the infinite-temperature state. Conversely, whenever the two operators have one or more eigenvectors in common in their spectral decomposition, the state of the monitored system converges differently in the subspaces spanned by the measurement observable eigenstates. As a result, we show that the convergence does not tend to a completely mixed (infinite-temperature) state, but to a block-diagonal state in the observable basis, with a finite effective temperature in each measurement subspace. Finally, we quantify the effects of the quantum hardware noise on the data by modelling them by means of depolarizing quantum channels. 
\end{abstract}

\maketitle

\section{Introduction}
In a quantum dynamical system, the application of a measurement necessarily entails a back-action on the system dynamics. Specifically, if one takes into account the von-Neumann postulate of quantum measurement (that for all practical purposes is just an idealization), after the measurement of a quantum observable, the quantum system instantaneously collapses onto an eigenstate of the observable. Such a mechanism is the well-known projective measurement~\cite{SakuraiBook}.

Quantum monitoring emphasizes the quantum measurement postulate since it prescribes the application over time of a sequence of projective measurements in single realizations of quantum system dynamics~\cite{Jacobs2006CM,JacobsBook}. In correspondence of each projective measurement, the quantum system collapses and re-starts its evolution from one of the quantum observable eigenstates with a given probability. This entails the onset of an ensemble of quantum system trajectories; the dimension of such an ensemble grows as $d^{n}$ with $d$ dimension of the quantum system and $n$ number of projective measurements applied in the single dynamical process.

In the last decade, several research groups in the quantum thermodynamics community have studied the property and effects of fluctuations due to measurement back-action~\cite{Elouard2017npjQI} in a sequence of quantum measurements from the point of view of thermodynamic quantities (work, heat, entropy) evaluated at least over two times~\cite{CampisiPRL2010,Campisi2011PRE,LeggioPRA2013,YiPRE2013,Horowitz2013NJP,WatanabePRE2014,GherardiniPRE2018,Mancino2018npjQI,MartinsPRA2019,Giachetti2020cm,Hernandez-GomezPRR2020,RossiPRL2020,MillerPRE2021,Hernandez-GomezPRXQuantum2022,Hernandez-GomezArxiv2022}. For more details, we refer the reader to a couple of recent reviews on the relation between quantum thermodynamics and continuous quantum monitoring~\cite{Manzano2022AVSQ} from the one hand, and repeated quantum measurements~\cite{Gherardini2022CSF}, corresponding to the discrete case, from the other hand. 

Another research line around quantum monitoring is the investigation of the behavior of the monitored system, both in the asymptotic limit of a large number of measurements~\cite{Talkner2011PRA,Gherardini2021PRE} and in the classical limit of the quantum system's size approaching macroscopicity~\cite{Gherardini2021PRE}. The tendency of the monitored quantum system to converge towards a steady state has been analyzed also in the more general case in which an additional source of noise affects the evolution between consecutive measurements~\cite{GherardiniNJP2016stochastic,GherardiniPRA2019Exact,Mueller2020PLA,ZieglerJSTAT2021,DasJSTAT2022,DidiPRE2022}. Specifically, in agreement with Refs.~\cite{HatridgeScience2013,PiacentiniNatPhys2017}, it has been determined that for any quantum observable the probability distribution of its outcomes, in the single realization of the system evolution, has an exact large-deviation form with an exponentially decaying profile in the number of measurements. This has allowed the derivation of the \emph{most probable distribution} of the observable outcomes~\cite{GherardiniPRA2019Exact}.

Furthermore, in this framework, it is worth mentioning also studies on quantum monitoring in the condensed matter context. It has been shown, indeed, that in quantum many-body systems the quantum measurement back-action in a sequence of measurements locally acting on the system is responsible for spontaneous symmetry breaking~\cite{Garcia-PintosPRL2019}, as well as for entangling-disentangling transitions~\cite{SkinnerPRX2019,RossiniPRB2020,Xhek2021PRB,Xhek2022PRB,Coppola2022PRB} recently experimentally observed~\cite{koh2022experimental}.

Despite this broad theoretical interest in quantum monitored systems, there is still a lack of benchmarks for many of the theoretical results previously summarized. At the same time, quantum hardware are emerging as versatile and easily accessible platforms~\cite{Tacchino2020AQT}. Thus, they can validate a large number of theoretical results, which were not (or very little) corroborated so far. In the last years, among these, a plethora of studies has been conducted on such devices~\cite{mi2022time,Frey2022SA,Francis2021SA,TornowArxiv2022,Santini2022PRA,LesovikSciRep2019,BuffoniPRL2022Third,Cattaneo2022ArXiv,Quintero2022FQST,Keenan2022ArXiv,Satzinger2021Science,Mi2022ArXiv}. In particular, in the context of quantum thermodynamics, several experiments have been performed on quantum hardware to benchmark quantum fluctuation relations~\cite{Solfanelli2021PRXQ,GherardiniPRAend-point}, to test their robustness against intermediate projective measurements~\cite{Talkner2011PRA}, and to experimentally realize~\cite{Solfanelli2021PRXQ,Solfanelli2022AVSQ,GardasSciRep2018,buffoni2020QST,CampisiPRE2021} quantum heat engines and refrigerators~\cite{heat_engine1,heat_engine2,heat_engine3,buffoni2019quantum,solfanelli2019maximal}. All these experiments are aimed of characterizing the energetic costs of quantum computation~\cite{Montangero2022GreenQuantumAdvantage,Auffeves2022PRXQuantum,Stevens2022PRL,Cimini2020npjQI,Myers2022AVSQ,GiananiArxiv2022diagnostics}.

We thus wonder: may quantum monitoring be effectively applied on a (commercial) quantum device? May such devices be exploited to investigate the open research issues under quantum monitoring, especially in the limit of a large number of measurements? In this paper, on the IBM Quantum hardware, encoding single-qubit and two-qubit systems, we demonstrate the phenomena of partial and infinite-temperature thermalization~\cite{Giachetti2020cm,Gherardini2021PRE} in the subspaces spanned by the measurement quantum observable. It is important to mention some related experimental~\cite{referee1_1,referee1_2,referee1_3,referee1_4,referee1_5,referee1_6,referee1_7} and theoretical~\cite{referee1_8,referee1_9} works that deal with the interplay between Hamiltonian evolution
and measurements. The properties of the hardware, the fabrication of the superconducting circuit and the methodology of the measurement scheme can be found on the IBM Quantum notes~\cite{IBMQ_ref}.

In this paper, we study the interplay between the unitary dynamics induced by a time-independent Hamiltonian $H=\sum_k E_k \dyad{E_k}$ and repeated projective measurements of a measurement observable $\mathcal{O} = \sum_k o_k \dyad{\alpha_k}$. Specifically, we consider a protocol that consists of repeated cycles from $n$ to $n+1$, where each cycle involves a unitary evolution up to a time $\tau$ followed by a projective measurement of the system in the basis of $\mathcal{O}$.

\emph{Infinite-temperature thermalization} (ITT) denotes the tendency of a monitored quantum system to end up towards the completely mixed state in the limit of a large (ideally infinite) number $n$ of measurements. Such behavior complies with the fact that the dynamics originating by a sequence of projective measurements interspersed with a unitary evolution is described by a \emph{unital} quantum map. However, as determined in Ref.~\cite{Gherardini2021PRE}, ITT occurs under some conditions: if {\rm (i)} the Hamiltonian $H$ of the system is a block matrix in the eigenvectors basis of the measurement observable $\mathcal{O}$, and {\rm (ii)} the size of the monitored quantum system is finite. Conversely, in case $H$ and $\mathcal{O}$ share a common nontrivial subspace (it is indeed enough that they are commuting, or even nearly commuting, operators), the so-called \emph{partial thermalization} of the system's state is recovered in the limit of $n\rightarrow\infty$. Partial thermalization manifests itself in the fact that the Hamiltonian $H$, once expressed in the basis decomposing the measurement observable $\mathcal{O}$, is a block matrix and, at the same time, the largest eigenvalue of the matrix containing the transition probabilities to jump from an eigenstate of $\mathcal{O}$ to another in single trajectories is degenerate. As a result, the monitored quantum system converges to a block-diagonal density operator in the quantum observable basis that is characterized by a finite effective temperature in each measurement subspace. Also, the period $\tau$ between measurements plays a role. First of all, as long as $n\tau \ll 1$ and the monitored system is initialized in an eigenstate of the measurement observable, the system dynamics is `frozen' in the initial condition in agreement with the so-called \emph{quantum Zeno effect}~\cite{ItanoPRA1990,FacchiJPAMT2008}. Secondly, one can observe \emph{resonance}-like behaviors when the energy gap of the monitored system is commensurable with the inverse of $\tau$. This entails that, for fixed values of $\tau$, the system jumps from an eigenstate of $\mathcal{O}$ to another as a function of $n$, integer number. Such resonance-like behavior is delectable in the expectation value of $\mathcal{O}$. Finally, due to the hard constraints imposed on the number of measurements by the noise in the quantum hardware, we model its (dynamical) effect on the measured statistics as a function of $n$. Specifically, we determine that, for single-qubit and two-qubit systems, ITT and partial thermalization are disturbed by a noise process modeled by a depolarizing quantum channel, which might have originated from using controlled-NOT gates. Despite the presence of noise hindering us to apply an arbitrarily large number of quantum measurements in a single trajectory, our experiments may help to understand the interplay of unitary dynamics interspersed by projective measurements in contact with an environment leading to decoherence. 

\section{Quantum monitoring}
\begin{figure}
    \centering
    \includegraphics[width=.8\linewidth]{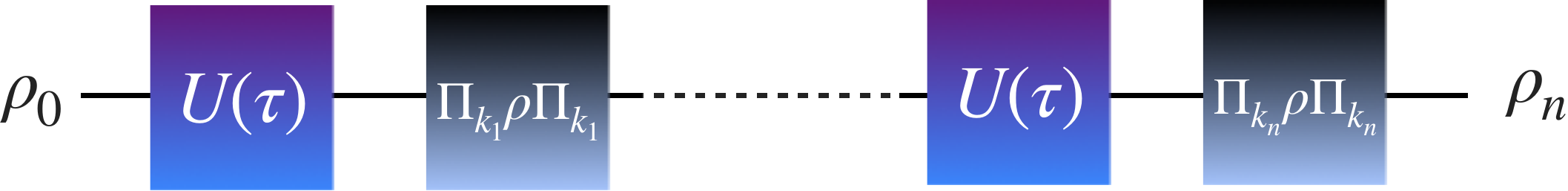}
    \caption{
    Pictorial representation of the quantum monitoring protocol, implemented on the IBM quantum hardware. 
    }
    \label{fig:figure0}
\end{figure}
Let us briefly recall the main theoretical results for the protocol we are going to take into account~\cite{Gherardini2021PRE}. To this extent, we consider a quantum system that is defined in an $N$-dimensional Hilbert space and evolves with a time-independent Hamiltonian $H$. At time $t = 0$, the system is initialized in the pure state $\rho_0 \equiv \dyad{\Psi_0}$ (in Ref.~\cite{Gherardini2021PRE}, $\ket{\Psi_0}$ is an eigenstate of the Hamiltonian, obtained by performing a first projective measurement of the energy). At times $t_{n} \equiv n\tau$, with $n$ integer number, the system is subjected to a sequence of projective measurements of a generic observable $\mathcal{O}=\sum_{k}o_{k}\Pi_{k}$. In general, $[H,\mathcal{O}] \neq 0$. Between two consecutive measurements, the monitored system follows the unitary evolution $U \equiv \exp(-iHt/\hbar)$, where $\hbar$ is the reduced Planck constant that is set to $1$ from now on. Then, after each projective measurement the system collapses in one of the eigenstates, $\ket{\alpha_k}$, of $\mathcal{O}$ with probability $P_{\ket{\alpha_k}} = \Tr[\rho\Pi_k]$. The protocol is sketched in Fig.~\ref{fig:figure0}. Notice that, if $[H,\mathcal{O}] \neq 0$, $\lbrace\ket{\alpha_k}\rbrace$ are not eigenstates decomposing $H$. This means that, in a sequence of projective measurements of $\mathcal{O}$ interspersed by the unitary evolution $U$, the system dynamics 
\begin{equation}
    \rho_n = \sum_{k_{t_n}, \ldots, k_{t_1}} \Pi_{k_{t_n}} U \ldots \Pi_{k_{t_1}} U \rho_0 \, U^{\dagger} \Pi_{k_{t_1}} \ldots U^{\dagger}\Pi_{k_{t_n}}
\end{equation}
originates, in a not trivial way, a bunch of quantum trajectories as well as a multi-time statistics for the outcomes of $\mathcal{O}$. In particular, the transition probability between two eigenstates of $\mathcal{O}$ at two consecutive times is given by the matrix element
\begin{equation}\label{Lmatrix}
    L_{k,k^{\prime}} \equiv  \abs{\matrixelement{{\alpha_{k^{\prime}}}}{U(\tau)}{{\alpha_{k}}}}^2 \,.
\end{equation}
As a consequence, the state of the system at time $t= n \tau$ is equal to the statistical mixture 
\begin{equation}
     \rho_n = \sum_k P^n_{\ket{\alpha_k}} \dyad{\alpha_k},
\end{equation}
where the probabilities $P^n_{\ket{\alpha_k}}$ are given by
\begin{equation}\label{Markov}
    P^n_{\ket{\alpha_k}} = \sum_{k^{\prime}} (L^{n})_{k, k^{\prime}} P^{0}_{\ket{\alpha_{k^{\prime}}}} 
\end{equation}
with $P^{0}_{\ket{\alpha_k}} \equiv |\braket{\alpha_k}{\Psi_0}|^2$.

The occurrence of the observable outcomes (at discrete times), in the single trajectory, is effectively described by a Markov process. The condition ${\rm Tr}[\rho_n] = 1$, ensuring the normalization of the probabilities, is preserved at any time by the fact that matrix $L$ is stochastic, namely   
\begin{equation}
    \sum_{k^{\prime}} L_{k,k^{\prime}} = 1 
\end{equation}
$\forall \ k^{\prime} = 1, \dots, N$.
The behavior of the system dynamics at large times ($n$ large) can be inferred by looking at the spectrum of $L$, as proved in Ref.~\cite{gardiner1994handbook}. As $L(\tau)$ is symmetric and stochastic, it follows that the spectrum $\lbrace \lambda_k \rbrace$ of $L$ is real and such that $-1 \leq \lambda_k \leq 1$ $\forall k$, while the largest eigenvalue $\lambda_0$ is $=1$. In the $\lbrace \ket{\alpha_k} \rbrace$ basis, it corresponds to the eigenvector
\begin{equation}
    \ket{\mathbf{v}_0} = \frac{1}{\sqrt{N}} (1, \dots, 1)^{T}.
\end{equation}
As shown in Ref.~\cite{Gherardini2021PRE}, if $H$ is irreducible in the basis $\lbrace \ket{\alpha_k} \rbrace$, the Perron-Frobenius theorem guarantees that $\lambda_0$ is non-degenerate. Thus, for large $n$, denoting with $\rightarrow$ the limit of $n \to \infty$, we obtain
\begin{equation}
    L^{n} \rightarrow \dyad{\mathbf{v}_0}
\end{equation}
so that $P^n_{k} \rightarrow 1/N$ and the monitored system evolves towards an infinite-temperature (i.e., completely mixed) state. This case is referred to as infinite-temperature thermalization~\cite{Giachetti2020cm,Gherardini2021PRE}. However, if $\mel{\alpha_k}{H}{\alpha_{k^{\prime}}}$ is a block matrix, then the eigenvalue $\lambda_0 = 1$ is degenerate, entailing that the dynamics will flow towards a state that conserves the memory of the initial condition. This case is referred to as partial thermalization~\cite{Gherardini2021PRE}. Another special case where the thermalization is prevented is when there is an eigenvalue $\lambda=-1$. In this case, the limit $P^n_k$ for $n\rightarrow \infty$ does not exist, as the system oscillates from the eigenvectors of the measurement observable. For small values of $n$, this phenomenon may occur also for particular values of $\tau$, due to the presence of resonances between commensurate characteristic frequencies in the level spacing of $H$. 

In the following, we will demonstrate the predictions of our theoretical framework on the quantum hardware.

\section{Single-Qubit}
\label{sec:III}

In order to demonstrate the aforementioned dynamics, we begin by considering a single-qubit $q_0$. The Hilbert space of the corresponding two-level system is described by the two eigenstates of $\sigma_{q_0}^z=\dyad{0}-\dyad{1}$, commonly known as the computational basis $\ket{k}$ with $k=0,1$. The starting point of our protocol is the state $\ket{\Psi_0} = \ket{0}$, which undergoes a cyclic dynamics given by an evolution governed by the following Hamiltonian:
\begin{equation}
     H_{q_0} = \frac{1}{2}\sigma_{q_0}^x,
\end{equation}
up to a time $\tau$ followed by a projective measurement in the $\sigma_{q_0}^z$ basis. 

We are going to monitor the `magnetization' along $z$ of the qubit that is given by the imbalance of its populations and reads as
\begin{equation}\label{eq:AverageMagnetization}
    \left<\sigma^z_{q_0}\right>(n \tau) = P^n_{\ket{0}}-P^n_{\ket{1}} \,. 
\end{equation}

\begin{figure}
    \centering
    \includegraphics[width=\linewidth]{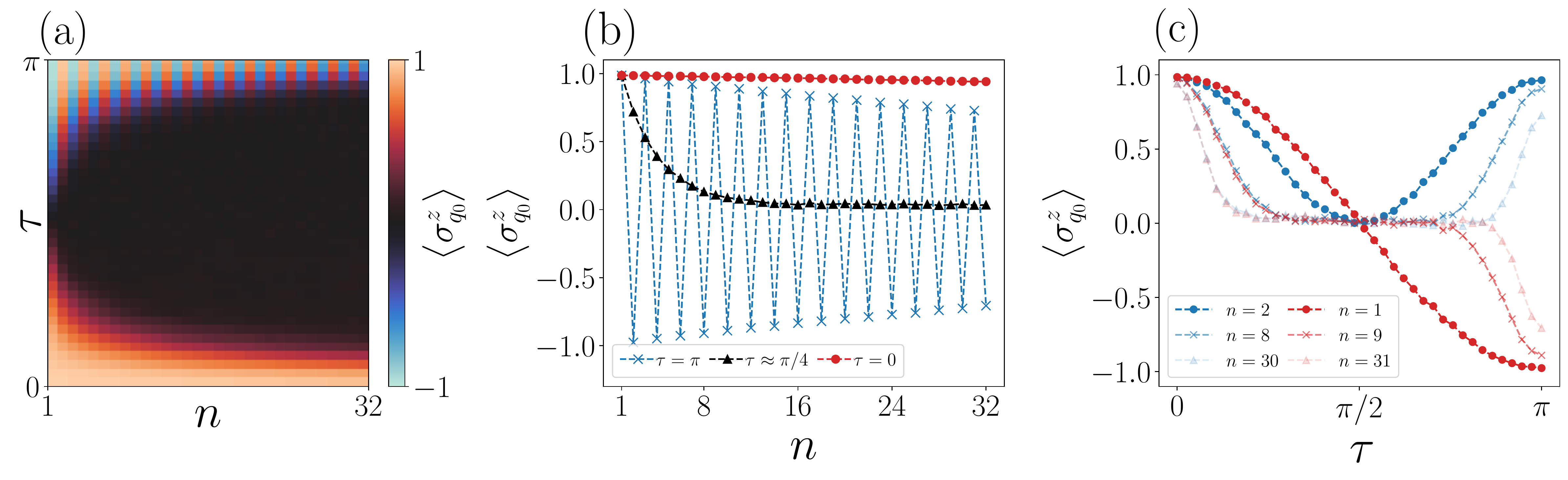}
    \caption{
    Observed values of $\left<\sigma^z_{q_0}\right>(n \tau)$ as a function of $n\in[1,32]$ and $\tau\in[0,\pi]$, for single-qubit rotations around $x$ ($ H_{q_0} = \frac{1}{2}\sigma_{q_0}^x$) under repeated quantum measurements of its population along $z$. The statistical error, which is about $10^{-2}$, is smaller than the marker size. The circuit depth after the transposition into the basis of the gates is $253$. We estimate that the circuit duration from $n$ to $n+1$ is around $0.09\ \mu s$.
    }
    \label{fig:figure1}
\end{figure}

While in this case $P^0_{\ket{0}} = 1$ and $P^0_{\ket{1}} = 0$ ($\ket{\Psi_0} = \ket{0}$ indeed), in order to compute the probabilities $P^{n}_{\ket{k}}$ ($k=1,2$) at later times we need the transition matrix $L$ defined in Eq.~\eqref{Lmatrix}. In the single-qubit case, the transition matrix explicitly reads as
\begin{equation}
    L = \begin{pmatrix*}[r]
    \cos^2 \left(\frac{\tau}{2}\right) & \sin^2 \left(\frac{\tau}{2}\right) \\ 
    \sin^2 \left(\frac{\tau}{2}\right) & \cos^2 \left(\frac{\tau}{2}\right)
    \end{pmatrix*}.
\end{equation}
Then, the spectral decomposition of $L(\tau)$ allows us to compute
\begin{equation}
    L^n = \lambda_0^n \dyad{\mathbf{v}_0} + \lambda_1^n \dyad{\mathbf{v}_1},
\end{equation}
with
\begin{align}
    \lambda_0 = 1 && &\ket{\mathbf{v}_0} = \frac{1}{2}\begin{pmatrix*}
    1 && 1
    \end{pmatrix*}^T,\notag\\
    \lambda_1 = \cos \tau && &\ket{\mathbf{v}_1} = \frac{1}{2}\begin{pmatrix*}
    1 && -1
    \end{pmatrix*}^T. 
\end{align}
As a consequence, following Eq.~\eqref{Markov} and Eq.~\eqref{eq:AverageMagnetization} we can compute the `magnetization' along $z$ of the qubit
\begin{equation}\label{eq:cos_t_n}
    \left<\sigma^z_{q_0}\right>(n\tau) = \left(\cos\tau\right)^n 
\end{equation}
that, in most cases, relaxes to the infinite-temperature thermal expectation value, namely $\left<\sigma_z\right> = 0$. However, as already noticed, exceptions to this general behavior do arise for some particular choices of $\tau$, i.e., $\tau  = p\pi$ with $p \in \mathbb{Z}$, corresponding to resonance-like behaviors. This is clear from Figure~\ref{fig:figure1}(a), where the observed data for $\left<\sigma^z_{q_0}\right>$ --obtained from the protocol implementation on a real quantum processor-- are shown. 
ITT is achieved for most of evolution times $\tau$ except for a set of points near the fine-tuned cases $\tau = 0,\pi$. On top of this behavior, the resonance can be clearly identified. When $\tau = 0$ then the system is frozen in the initial state $\ket{\Psi_0} = \ket{0}$ with $\left<\sigma^z_{q_0}\right> = 1$, see Fig.~\ref{fig:figure1}(b) red circles. Close to this value of $\tau$, the relaxation timescale is $n \sim O(\tau^{-2})$ so that the dynamics of the monitored system is frozen within it. This is nothing but a manifestation of the quantum Zeno effect  that manifests itself for small but nonzero $\tau$. This means that it takes $O(\tau^{-2})$ steps to populate the state $\ket{1}$ in the presence of measurements, thus preventing thermalization as a consequence of repeated collapses of the system on the initial state. 
Moreover, whenever $\tau = \pi$, we find a second resonance effect, as $\left<\sigma^z_{q_0}\right> = (-1)^n$ (see Eq.~\eqref{eq:cos_t_n}). The latter stems from taking, at $\tau = \pi$, the energy gap of the qubit \emph{commensurable} with the inverse of the period, as shown in Fig.~\ref{fig:figure1}(b) blue crosses. In other words, in this case, the unitary evolution simply brings the system back and forth between the eigenstates of the measured observable, corresponding to $\left<\sigma^z_{q_0}\right> = \pm 1$, respectively. In this case, the thermalization timescale goes as $n = O((\pi-\tau)^{-2})$ as well. Interestingly, this mechanism is analogous to the phenomenon of \emph{period doubling} of Floquet driven systems near to resonances~\cite{Wilczek2012PRL,Khemani2016PRL,Else2020Review,zhang2017observation,choi2017observation,Santini2022PRB,Collura2022PRX,Fazio2017PRB,Fazio2019PRB,Fazio2021PRB,pizzi2021higher,Giachetti2022Arxiv}. 

Such effects are well highlighted, albeit in a different way, also by Fig.~\ref{fig:figure1}(c) where we fix the value of $n$ (in the figure, $n=1,2,8,9,30,31$) and $\left<\sigma^z_{q_0}\right>$ is plotted as a function of $\tau\in[0,\pi]$. In the sub-interval $\tau\in[\pi/2,\pi]$, we observe that the monitored system (starting from $\ket{\Psi_0} = \ket{0}$) tends to flip towards the excited state $\ket{1}$ or to stay in the ground $\ket{0}$, depending on the fact that $n$ --integer number-- is even ($\left<\sigma^z_{q_0}\right> > 0$) or odd ($\left<\sigma^z_{q_0}\right> < 0$). Moreover, these behaviors occur in a more discontinuous way the greater the number of projective measurements. A discontinuity is found in the limit of $n\rightarrow\infty$, whereby in such a limit resonances are events of measure zero (i.e., occurring with zero probability almost surely).

In practice, to implement a quantum circuit realizing our protocol on the quantum processor we need to decompose it into the set of native gates, which are device dependent~\cite{IBMQ_ref}. This procedure and its optimization are referred to as transpiling the circuits and can be automatically implemented with Qiskit~\cite{Qiskit-Textbook}. We can evaluate the actual execution time of the transpiled circuit into the quantum hardware native gates thanks to the properties of the hardware that we show in~\ref{app:Hardware} in table~\ref{tab:hardware_properties}. To achieve this, one counts the number of layers for each operation involved in the circuit. Such number of layers, indeed, corresponds to the number of non-parallelizable operations that contribute to the circuit's depth. In particular, we find that to implement an unitary evolution and projective measurement, approximately $7$ layers of single-qubit gates and one layer of measurement are needed for the evolution from $n$ to $n+1$, which implies an evolution time of about $\Delta t = 0.09\ \mu s$. Notice that the number of gates applied in each cycle from $n$ to $n + 1$ is kept constant. This entails that also the time elapsed on the quantum hardware for a cycle is approximately constant. Thus, the estimate of the circuit duration $n\Delta t$ refers to the average time $\Delta t$ in a single cycle, regardless of the value of $\tau$.

Finally, we remark that each evolution, together with the needed measurements, is repeated $n_\mathrm{shots} = 2^{13} = 8192$ times, so that the statistical error on the measurement outcomes could be estimated to be of the order of $1/\sqrt{n_\mathrm{shots}} \approx 10^{-2}$. However the corresponding error bars are not shown in our figures since they are smaller than the marker size.

\section{Partial thermalization of two qubits}\label{sec:IV}

We consider a composite system of two non-interacting qubits $q_0$, $q_1$ whose dynamics is described by the Hamiltonian
\begin{equation}
    H = \frac{1}{2}\left(\sigma_{q_0}^x+ \sigma_{q_1}^x\right)\label{eq:two_qubit_ham}
\end{equation}
and, at time $t=0^{-}$, starts from the product state $\rho_0=\dyad{00}{00}$. Our protocol is based on applying a sequence of evolutions intertwined by projective measurements of an observable whose eigenbasis (spectral decomposition) contains entangled states with respect to the computational basis. In this regard, the change of basis from a generic two-qubit basis $\{\ket{\phi_k}\}$, with $k=0,1,2,3$, to the computational basis $\{\ket{k}\}$, with $k=00,01,10,11$, is given by the following relation:
\begin{equation}
    \begin{pmatrix}
    \ket{00}\\ \ket{01}\\ \ket{10}\\ \ket{11}
    \end{pmatrix} = V
    \begin{pmatrix}
    \ket{\phi_0}\\ \ket{\phi_1}\\ \ket{\phi_2}\\ \ket{\phi_3}
    \end{pmatrix},
\end{equation}
where $V$ is the matrix for the change of basis. We notice that, while the evolution acts locally and independently on the two qubits (due to our choice of the Hamiltonian $H$), the projective measurements may create entanglement as they allow for an effective exchange of information. However, an equivalent protocol could be achieved (with the same circuit complexity) by choosing an entangling dynamics and taking $\sigma^z_{q_j}$ as local observables whose spectral decomposition is the computational basis of the single-qubit. In fact, as a by-product of our protocol, we have determined that the same statistics of measurement outcomes can be recovered in the following two ways. {\rm (i)} By letting evolve a quantum system with local time-independent Hamiltonian and then (repeatedly) measuring it on a generic (entangling) basis. Or {\rm (ii)} by implementing dynamics under a generic entangling Hamiltonian and measuring the system in the local computational basis. Furthermore, we are also going to take the opportunity to show the way of realizing quantum measurements in a generic basis, decomposing a given quantum observable, on quantum hardware.

We stress that, on the IBM Quantum hardware~\cite{IBMQ_ref}, the only native measurement allowed is over the computational basis corresponding to the projectors $\Pi_k=\dyad{k}$. Thus, in order to perform measurements over the eigenprojectors of an entangled basis 
\begin{equation}
    \pi_k = \dyad{\phi_k},
\end{equation}
first we have to apply the change of matrix $V$ that maps the entangled basis $\{\ket{\phi_k}\}$ into the $\sigma^z$ eigenbasis. Then, we need to measure along $\sigma^z$, and finally, we have to perform the inverse unitary gate $V^\dagger$ that is implicitly defined by the relation
\begin{equation}
    \pi_k = V^{\dagger} \Pi_k V \,.
\end{equation}

In the following, our analysis will focus on two specific measurement basis, the single-triplet and the Bell bases, in which the conditions allowing partial thermalization are discussed.

\subsection{Singlet-Triplet basis}

The change-of-basis matrix $V$ that maps the singlet-triplet basis into the computational basis is given by
\begin{equation}
    \begin{pmatrix}\ket{00}\\ \ket{01}\\ \ket{10}\\ \ket{11}\end{pmatrix} =  \begin{pmatrix*}[r]
    1 & 0 & 0 &  0 \\
    0 & 1/\sqrt{2} & -1/\sqrt{2} & 0 \\
    0 & 1/\sqrt{2} & 1/\sqrt{2} & 0 \\
    0 & 0 & 0 & 1
    \end{pmatrix*}
    \begin{pmatrix}\ket{\psi_0}\\ \ket{\psi_1}\\ \ket{\psi_2}\\ \ket{\psi_3}\end{pmatrix},
\end{equation}
where the states\begin{equation}
    \ket{\psi_0} = \ket{00},\quad\ket{\psi_1} =\frac{1}{\sqrt{2}}\left(\ket{10}+\ket{01}\right),\quad\ket{\psi_4} = \ket{11},
\end{equation}
constitute the triplet state while
\begin{equation}\ket{\psi_2} =\frac{1}{\sqrt{2}}\left(\ket{10}-\ket{01}\right)
\end{equation}
is the singlet state. The latter is anti-symmetric under permutation $\mathcal{P}:q_0 \leftrightarrow  q_1$ of the two qubits, meanwhile the other three states are symmetric under such a transformation. 

Since the Hamiltonian~\eqref{eq:two_qubit_ham} is invariant under $\mathcal{P}$ and the initial state has a zero overlap with the subspace spanned by $|\psi_2\rangle$ (i.e., the anti-symmetric manifold), we therefore expect a partial thermalization in the symmetric manifold of the Hilbert space. In fact, in the absence of noise, the singlet state $\ket{\psi_2}$ is a dark state, and the Hamiltonian in the measurement basis reads as
\begin{equation}
    V^\dagger H V = \frac{1}{\sqrt{2}}\begin{pmatrix*}[r]
    0 & 1 & \vline\,0\,\vline & 0 \\
    1 & 0 & \vline\,0\,\vline & 1 \\
    \hline
    0 & 0 & \vline\,0\,\vline & 0 \\
    \hline
    0 & 1 & \vline\,0\,\vline & 0
    \end{pmatrix*}.
\end{equation}
Then, we may write the transition matrix $L(\tau)$ as 
\begin{equation}
L(\tau)=\begin{pmatrix*}[r]
\cos^4\frac{\tau}{2} & \frac{1}{2}\sin^2{\tau} & 0 & \sin^4 \frac{\tau}{2}\\
\frac{1}{2}\sin^2{\tau} & \cos^2 \tau & 0 & \frac{1}{2}\sin^2{\tau} \\ 0 & 0 & 1 & 0 \\
\sin^4 \frac{\tau}{2} & \frac{1}{2}\sin^2{\tau} & 0 & \cos^4\frac{\tau}{2}
\end{pmatrix*},
\end{equation}
and we have $P^0_{\ket{\psi_{k}}} = \delta_{k,0}$ due to the fact the initial state $\ket{\psi_0}$ is $\ket{00}$.
\begin{figure}
    \centering
    \includegraphics[width=.49\linewidth]{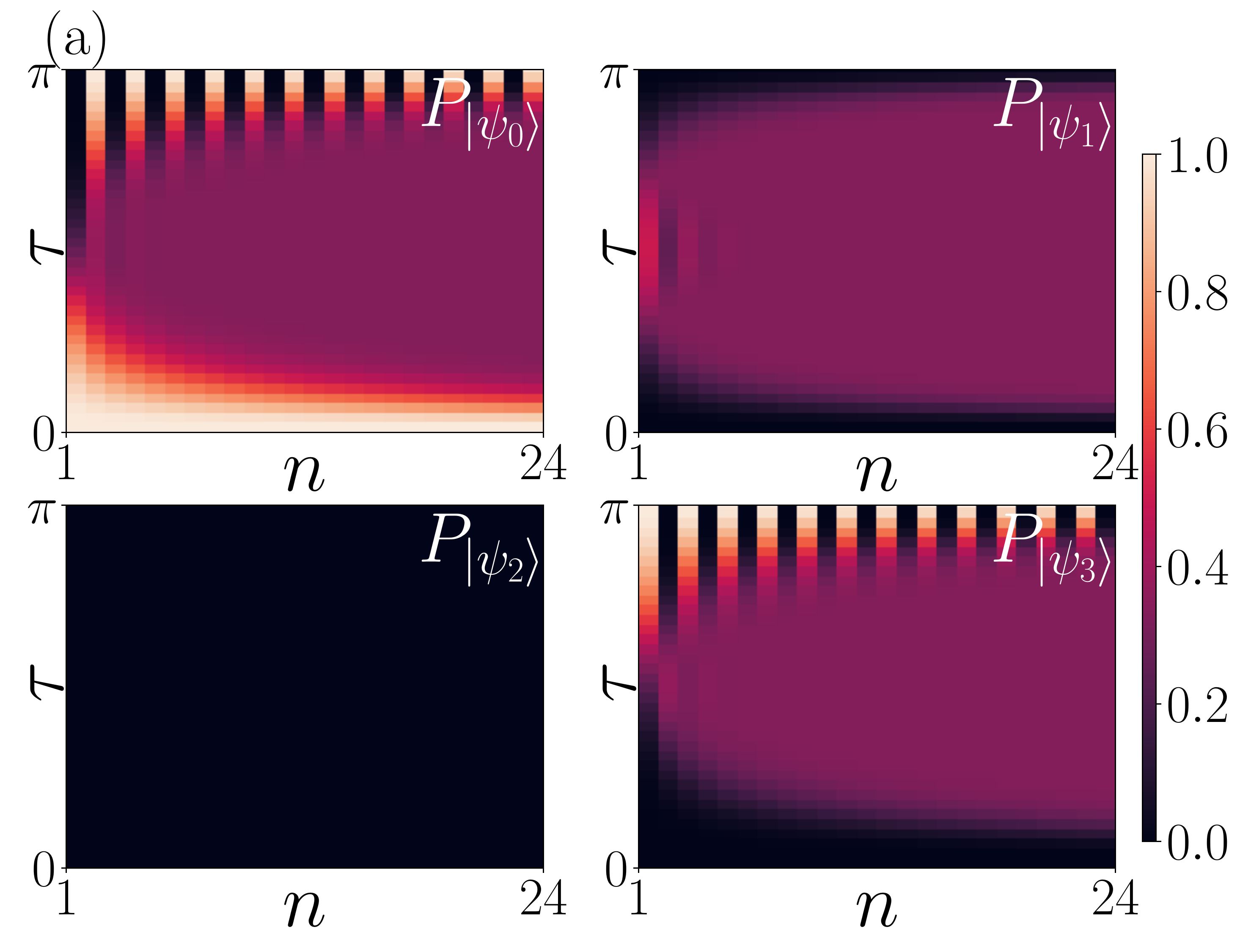}
    \includegraphics[width=.49\linewidth]{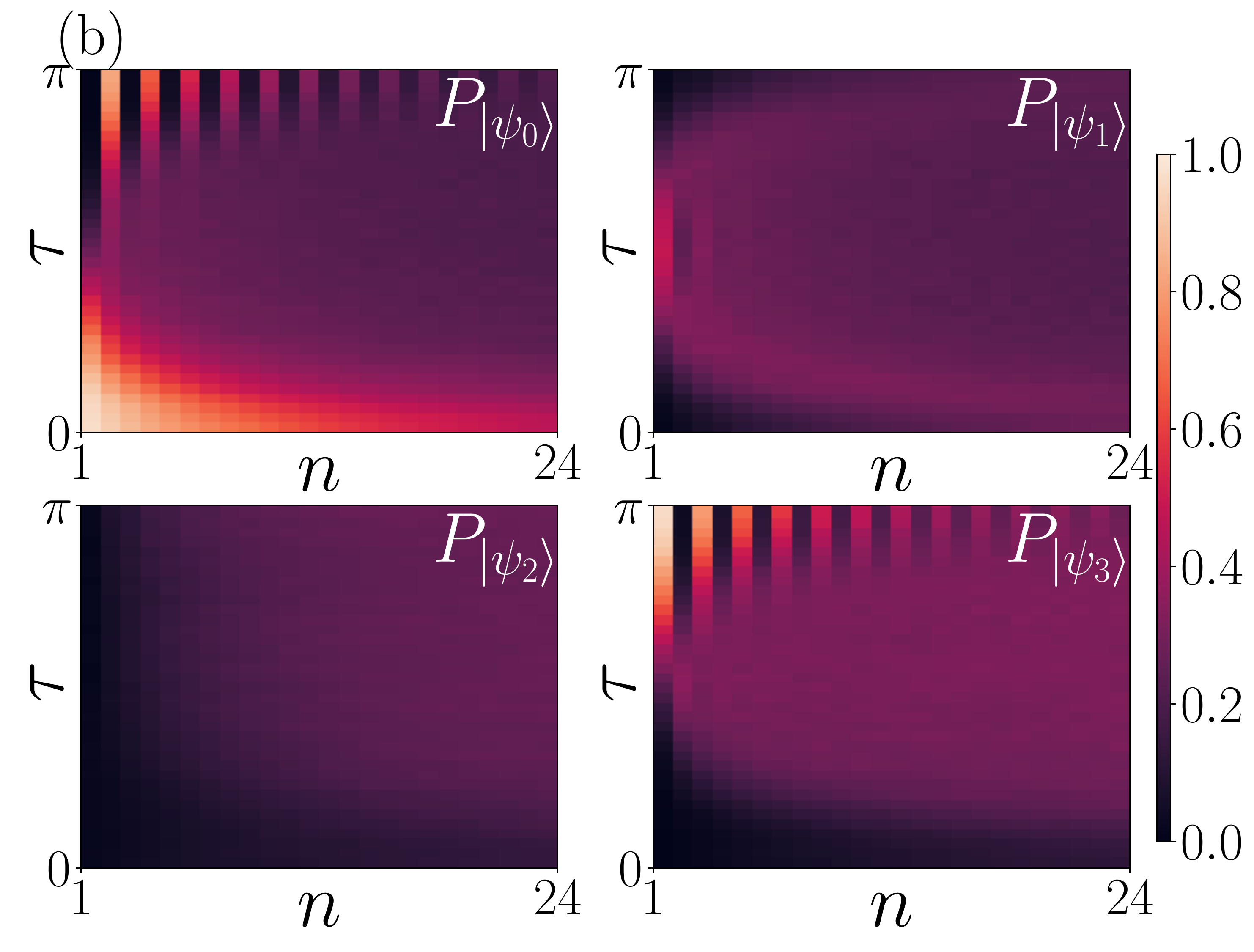}
    \caption{Theoretical, panel (a), and observed, panel (b), values of the probabilities to measure two rotating qubits, subjected to repeated quantum measurements, in the subspaces of the system Hilbert space spanned by the singlet-triplet basis $\{\ket{\psi_k}\}$, with $k=0,1,2,3$. The circuit depth after the transposition into the circuit basis gates is $577$. We estimate the circuit duration from $n$ to $n+1$ is of about $2.7\ \mu s$.}
    \label{fig:figure2}
\end{figure}
Let us now show $L^n(\tau)$, which is provided by the following expression:
\begin{equation}
    L^n(\tau) = \sum_{j=0}^3 \lambda^n_j \dyad{\mathbf{v}_j}
\end{equation}
where
\begin{align}
    \lambda_0 = 1, && &\ket{\mathbf{v}_0} = \frac{1}{\sqrt{3}}\begin{pmatrix*}[r]1 & 1 & 0 & 1 \end{pmatrix*}^T, \notag\\
    \lambda_1 = 1, && &\ket{\mathbf{v}_1} = \begin{pmatrix*}[r]0 & 0 & 1 & 0 \end{pmatrix*}^T,\notag \\
    \lambda_2 = \cos\tau, && &\ket{\mathbf{v}_2} = \frac{1}{\sqrt{2}}\begin{pmatrix*}[r]1 & 0 & 0 & -1 \end{pmatrix*}^T,\notag \\
    \lambda_3 = \frac{1}{4}\left(1+3\cos 2\tau\right), && &\ket{\mathbf{v}_3} = \frac{1}{\sqrt{6}}\begin{pmatrix*}[r]1 & -2 & 0 & 1 \end{pmatrix*}^T \,.
\end{align}
Thus, the probabilities $P^{n}_{\ket{\psi_k}} \equiv {\rm Tr}\left[\rho_{n}|\psi_k\rangle\!\langle\psi_k|\right]$ are
\begin{align}\label{eqs:theoprob_SingletTriplet}
&P^{n}_{\ket{\psi_0}} = \frac{1}{6} \left[3 \cos ^n(\tau )+2^{-2n} (3 \cos (2 \tau )+1)^n+2\right],\notag\\
&P^{n}_{\ket{\psi_1}} =\frac{1}{3} \left[1 - 2^{-2n} (3 \cos (2 \tau )+1)^n \right],\notag\\
&P^{n}_{\ket{\psi_2}}=0,\notag\\
&P^{n}_{\ket{\psi_3}}=\frac{1}{6}\left[-3 \cos ^n(\tau )+2^{-2n} (3 \cos (2 \tau )+1)^n+2\right],
\end{align}
which in the limit of $n\to \infty$ read
\begin{align}
&P^{\infty}_{\ket{\psi_0}} = \begin{cases}\frac{1}{3} \quad&\mathrm{if}\ \tau \neq p\pi,\ p \in \mathbb{Z},\\
1\quad &\mathrm{if}\ \tau = 2p\pi,\ p \in \mathbb{Z},\\
\frac{1}{2}\left((-1)^n+1\right)\quad &\mathrm{if}\ \tau=(2p+1)\pi,\ p\in \mathbb{Z},
\end{cases}\notag\\
&P^{\infty}_{\ket{\psi_1}} = \begin{cases}\frac{1}{3} \quad&\mathrm{if}\ \tau \neq p\pi,\ p \in \mathbb{Z},\\
0\quad &\mathrm{if}\ \tau = p\pi,\ p \in \mathbb{Z},
\end{cases}\notag\\
&P^{\infty}_{\ket{\psi_2}}=0,\notag\\
&P^{\infty}_{\ket{\psi_3}} = \begin{cases}\frac{1}{3} \quad&\mathrm{if}\ \tau \neq p\pi,\ p \in \mathbb{Z},\\
0\quad &\mathrm{if}\ \tau = 2p\pi,\ p \in \mathbb{Z},\\
\frac{1}{2}\left((-1)^{n+1}+1\right)\quad &\mathrm{if}\ \tau=(2p+1)\pi,\ p\in \mathbb{Z}\,.
\end{cases}
\end{align}

\begin{figure}
    \centering
    \includegraphics[width=.55\linewidth]{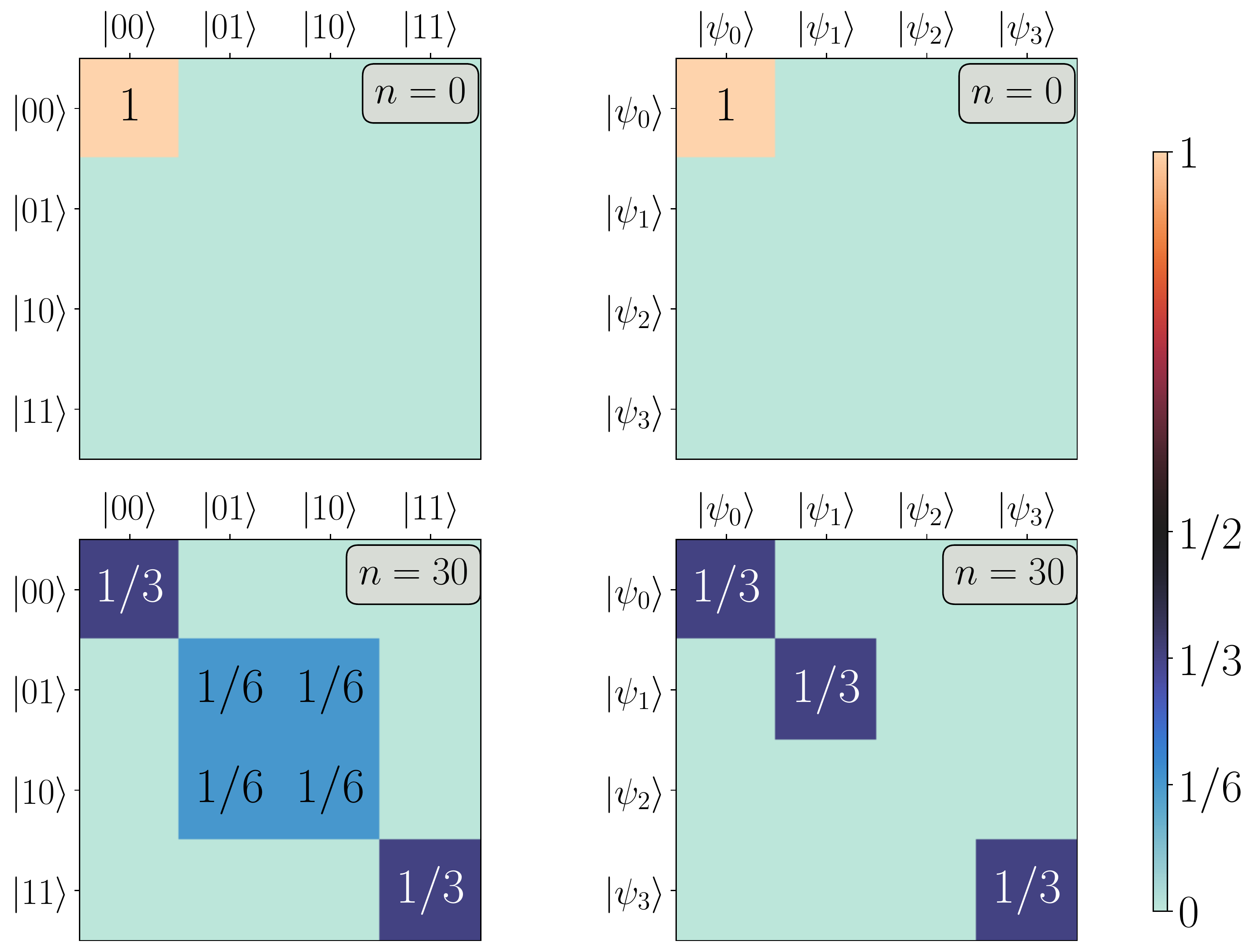}
    \caption{
    Elements of the theoretical density operators in the computational basis (panels on the left), and in the singlet-triplet basis (panels on the right), for $\tau=\pi/4$. The number of measurements is $n=0$ in the upper panels and $n=30$ in the lower ones.}
    \label{fig:figure7}
\end{figure}

The theoretical values of the probabilities in Eqs.~\eqref{eqs:theoprob_SingletTriplet}, plotted in Fig.~\ref{fig:figure2}(a), have to be compared with the ones obtained with the quantum hardware, which are shown in Fig.~\ref{fig:figure2}(b). Since the initial density operator $\rho_0 = \dyad{00}$ is also an eigenprojector of the measurement basis, we are able to observe the quantum Zeno effect for $\tau = 0$, as in the previous paragraph. For small $n$ and far from resonances, the state of the system quickly relaxes towards an equally populated mixed state in the triplet subspace $\{\ket{\psi_0},\ket{\psi_1},\ket{\psi_3}\}$. This corresponds to a partial thermalization of the monitored quantum system, since the subspaces of the singlet and triplet states are not mixed by the dynamics. In Fig.~\ref{fig:figure7} we plot the elements of the initial theoretical density operator and the ones obtained after $n=30$ projective measurements of our protocol. 
%
%
It's worth noting that, since the relaxation converges exponentially with $n$, the results obtained after 30 measurements are a quite good approximation of the ones taking into account the asymptotic values reached for $n\to\infty$. For large $n$, the non-trivial structure of the density operator in the computational basis takes into account the fact that the measurements protects the symmetric manifold of the Hilbert space by maintaining the system within it over time. However, in our experiments, by increasing the number of measurements, the noise brings the system to the infinite-temperature state, thus restoring the ITT regime. Notice that such effect is beyond the description of Fig.~\ref{fig:figure7} that describes the ideal, noiseless case. We refer the reader to Sec.~\ref{sec:V} for a detailed analysis of the noise affecting the quantum hardware that we employed.

Once again, the single-cycle evolution time can be computed, at least approximately. In particular, we find that to implement the unitary evolution and projective measurement on the quantum hardware, approximately $20$ layers of single-qubit gates, $4$ layers of CNOTs, and one layer of measurement are needed for the evolution from $n$ to $n+1$, which implies an evolution time of about $\Delta t = 2.7\ \mu s$.

\subsection{Bell basis}

\begin{figure}
    \includegraphics[width=.49\linewidth]{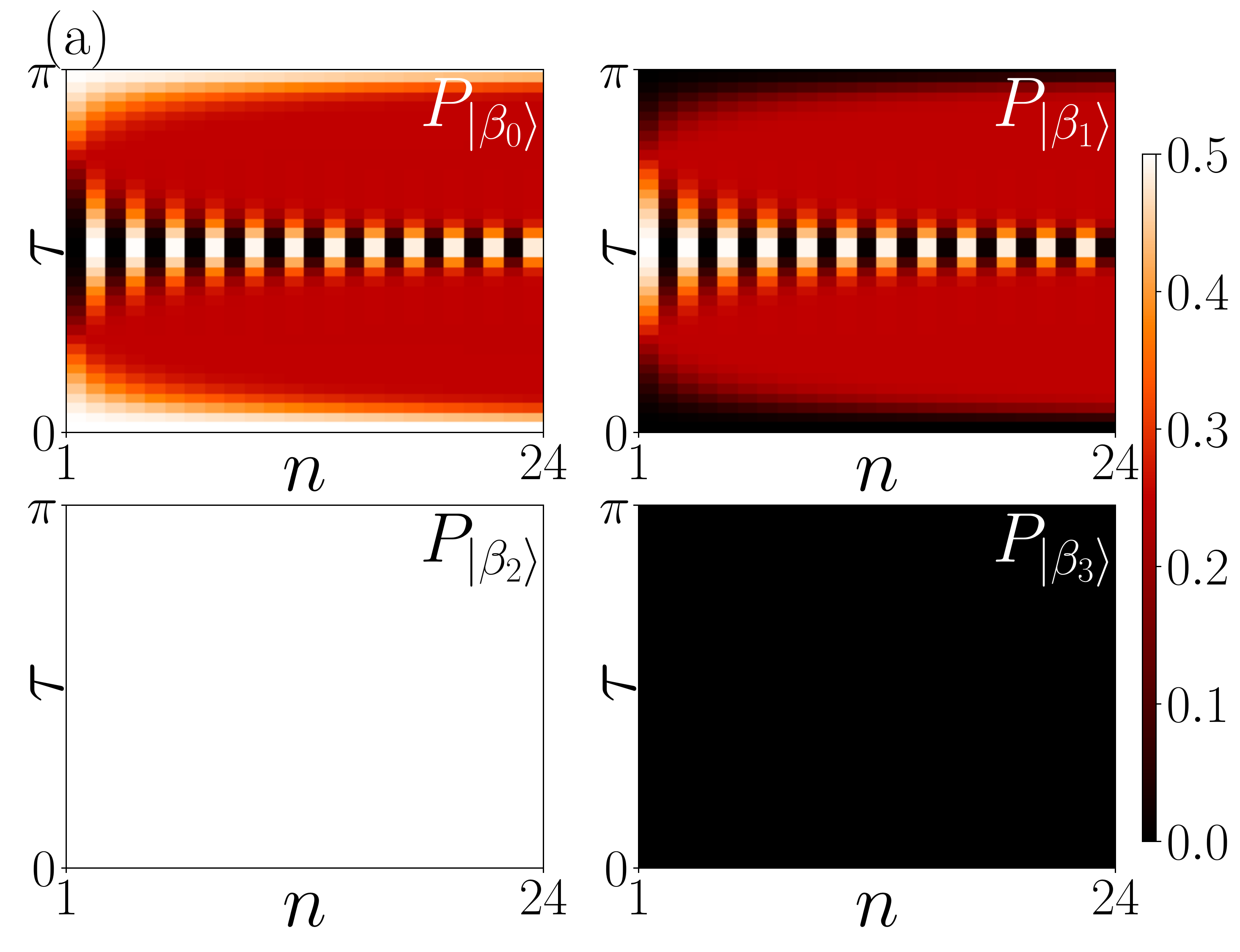}
    \includegraphics[width=.49\linewidth]{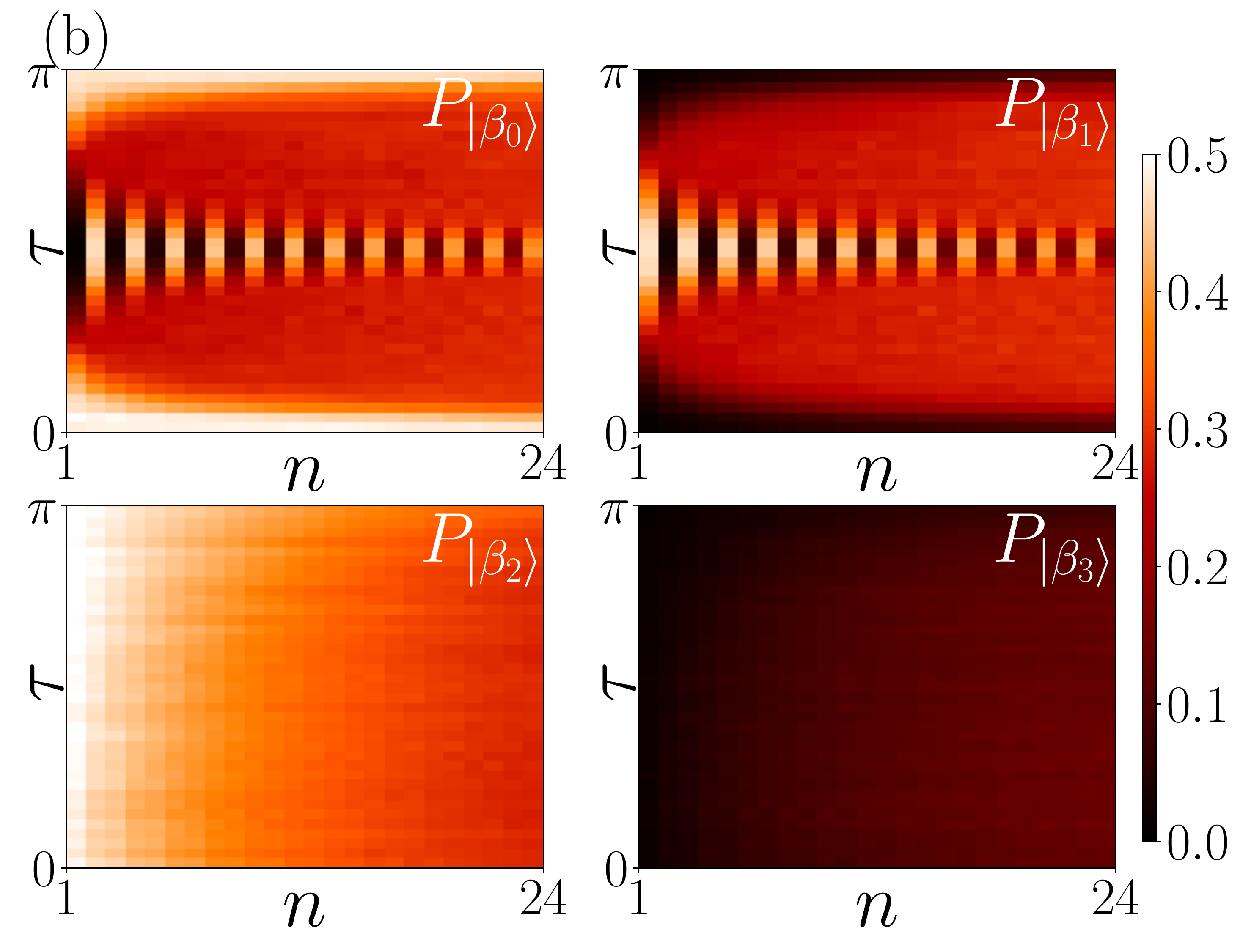}
    \caption{Theoretical, panel (a), and observed, panel (b), values of the probabilities to measure two rotating qubits, subjected to repeated quantum measurements, in the subspaces of the system Hilbert space spanned by the Bell basis $\{\ket{\beta_k}\}$, with $k=0,1,2,3$. The circuit depth after the transposition into the circuit basis gates is $312$. We estimate the circuit duration from $n$ to $n+1$ is of about $1.7\ \mu s$.}
    \label{fig:figure3}
\end{figure}
The second measurement basis that we have considered is the Bell basis. The change-of-basis matrix $V$ from the Bell states to the computational basis is
\begin{equation}
    \begin{pmatrix}\ket{00}\\ \ket{01}\\ \ket{10}\\ \ket{11}\end{pmatrix} = \frac{1}{\sqrt{2}} \begin{pmatrix*}[r]
    1 & 0 & 1 &  0 \\
    0 & 1 & 0 & 1 \\
    0 & 1 & 0 & -1 \\
    1 & 0 & -1 & 0
    \end{pmatrix*}
    \begin{pmatrix}\ket{\beta_0}\\ \ket{\beta_1}\\ \ket{\beta_2}\\ \ket{\beta_3}
    \end{pmatrix},
\end{equation}
from which we have
\begin{align}
    \ket{\beta_0} = \frac{1}{\sqrt{2}}\left(\ket{00} + \ket{11}\right),&& \ket{\beta_1} = \frac{1}{\sqrt{2}}\left(\ket{01}+\ket{10}\right),\notag\\
    \ket{\beta_2} = \frac{1}{\sqrt{2}}\left(\ket{00} - \ket{11}\right), && \ket{\beta_3} = \frac{1}{\sqrt{2}}\left(\ket{01}-\ket{10}\right).
\end{align}
Then, following the same procedure of the previous paragraph, the Hamiltonian in the measurement basis is
\begin{equation}
    V^\dagger H V = 
    \begin{pmatrix*}
    0 & 1 & \vline\,0 & 0\\
    1 & 0 & \vline\,0 & 0\\
    \hline
    0 & 0 & \vline\,0 & 0\\
    0 & 0 & \vline\,0 & 0\\
    \end{pmatrix*}.
\end{equation}
This means that in this case the initial state $\ket{00}$ is no longer an eigenstate of the measurement basis, so that 
\begin{equation}
P^0_{\ket{\beta_{k}}} = \frac{1}{2}\left( \delta_{k,0} +  \delta_{k,2}  \right).     
\end{equation}
Moreover, the transition matrix $L(\tau)$ is given here by
\begin{equation}
L(\tau)= \begin{pmatrix*}[r]
\cos^2(\tau) & \sin^2(\tau) & 0 & 0 \\
\sin^2(\tau) & \cos^2(\tau) & 0 & 0 \\
0 & 0 & 1 & 0 \\
0 & 0 & 0 & 1
\end{pmatrix*},
\end{equation}
with the result that
\begin{equation}
    L^n (\tau) = \sum_{j=0}^3 \lambda^n_j \dyad{\mathbf{v}_j}
\end{equation}
where 
\begin{align}
    \lambda_0 = 1, && &\ket{\mathbf{v}_0} = \begin{pmatrix*}[r] 0 & 0 & 0 & 1\end{pmatrix*}^T,\notag\\
    \lambda_1 = 1, && &\ket{\mathbf{v}_1} = \begin{pmatrix*}[r] 0 & 0 & 1 & 0\end{pmatrix*}^T,\notag\\
    \lambda_2 = 1, && &\ket{\mathbf{v}_2} = \frac{1}{\sqrt{2}}\begin{pmatrix*}[r] 1 & 1 & 0 & 0\end{pmatrix*}^T,\notag \\
    \lambda_3 = \cos 2\tau, && &\ket{\mathbf{v}_3} = \begin{pmatrix*}[r] 1 & -1 & 0 & 0\end{pmatrix*}^T.
\end{align}
It follows that $P^{n}_{\ket{\beta_k}}={\rm Tr}\left[\rho_{n}|\beta_k\rangle\!\langle\beta_k\right|]$ are 
\begin{align}\label{eqs:theoprob_Bell}
   &P^{n}_{\ket{\beta_0}} =  \frac{1}{4}\left(1+\cos ^n2 \tau \right),\notag &&
   P^{n}_{\ket{\beta_1}} =  \frac{1}{4}\left(1-\cos ^n2 \tau \right), \\
   &P^{n}_{\ket{\beta_2}} = \frac{1}{2}, &&
   P^{n}_{\ket{\beta_3}} = 0 \,,
\end{align}
which in the limit of $n\to\infty$ read
\begin{align}
    &P^{\infty}_{\ket{\beta_0}} = \begin{cases} \frac{1}{4} \quad&\mathrm{if}\ \tau \neq \frac{p}{2}\pi,\ p \in \mathbb{Z},\\
     \frac{1}{2}\quad&\mathrm{if}\ \tau = p\pi,\ p \in \mathbb{Z},\\
     \frac{1}{4}\left(1+(-1)^n\right)\quad&\mathrm{if}\ \tau = p\pi+\frac{\pi}{2},\ p \in \mathbb{Z},
    \end{cases}\notag\\
   &P^{\infty}_{\ket{\beta_1}} = \begin{cases} \frac{1}{4} \quad&\mathrm{if}\ \tau \neq \frac{p}{2}\pi,\ p \in \mathbb{Z},\\
     0\quad&\mathrm{if}\ \tau = p\pi,\ p \in \mathbb{Z},\\
     \frac{1}{4}\left(1+(-1)^{n+1}\right)\quad&\mathrm{if}\ \tau = p\pi+\frac{\pi}{2},\ p \in \mathbb{Z},
    \end{cases}\notag\\
   &P^{\infty}_{\ket{\beta_2}} = \frac{1}{2}, \notag\\
   &P^{\infty}_{\ket{\beta_3}} = 0 \,.
\end{align}

\begin{figure}
    \centering
    \includegraphics[width=.55\linewidth]{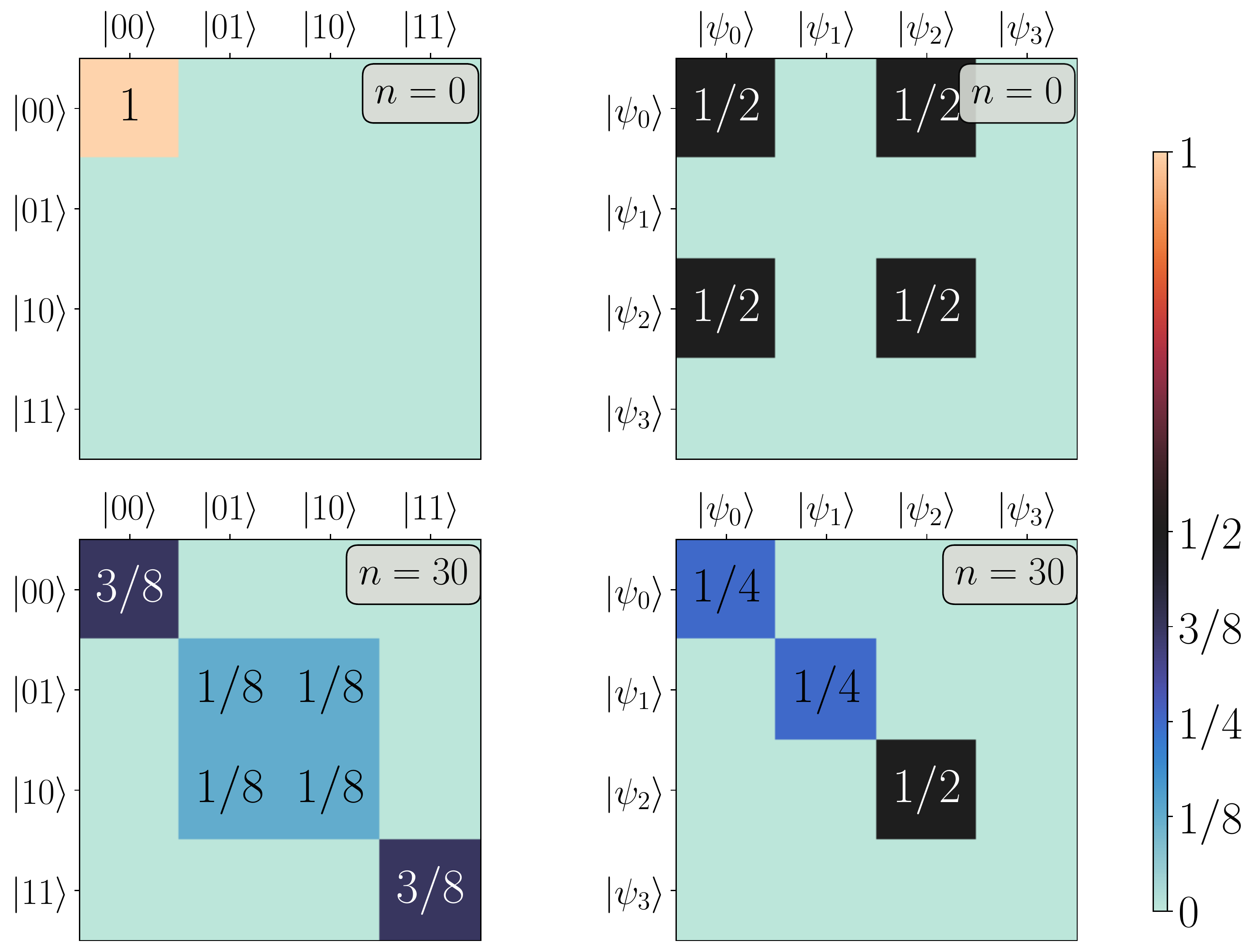}
    \caption{Elements of the theoretical density operators in the computational basis (panels on the left), and in the Bell measurement basis (panels on the right), for $\tau=\pi/4$. The number of measurements is $n=0$ in the upper panels and $n=30$ in the lower ones.}
    \label{fig:figure8}
\end{figure}

The theoretical populations of Eqs.~\eqref{eqs:theoprob_Bell}, plotted in Fig.~\ref{fig:figure3}(a), have to be compared with the ones obtained with the quantum hardware, shown in Fig.~\ref{fig:figure3}(b). In this case, the initial state of the dynamics, $\ket{00}=\frac{1}{\sqrt{2}}\left(\ket{\beta_0} + \ket{\beta_2}\right)$, is a superposition of two of the states of the measurement basis. Once again, at short time scales, partial thermalization is observed. In fact, $P^n_{\ket{\beta_3}} \approx 0$, $P^n_{\ket{\beta_2}} \approx 1/2$ $\forall n$ and only the complementary subspace is mixed by the dynamics. Moreover, there is less noise in comparison with the case-study in the previous paragraph. This could be due to the fact that a smaller number of CNOTs is needed to transpile the protocol into the native gates of the used quantum hardware. 

As in Fig.~\ref{fig:figure7}, we plot in Fig.~\ref{fig:figure8} the elements of the initial theoretical density operator and the ones obtained after $n=30$ projective measurements of our protocol. Once again, for large $n$, the non-trivial structure of the density operator in the computational basis allows for the protection of the symmetric manifold of the Hilbert space by maintaining the system within it over time.

As done in the previous paragraphs, we can estimate the evolution time of a single-cycle of the transpiled circuit. In the case of measuring over the Bell's basis, approximately $10$ layers of single-qubit gates, $2$ layers of CNOTs, and $1$ measurement layer have to be employed to reproduce the evolution from step $n$ to $n+1$. This implies an evolution time of about $\Delta t = 1.7\ \mu s$. We also note that, in this setting, the system dynamics is less noisy than in the case of the single-triplet measurement basis and the evolution faster, since $2$ fewer CNOTs are needed.

\section{Effects of the quantum hardware noise}\label{sec:V}

In our experiments, the simplest model able to describe the noise on the hardware is the global depolarizing channel
\begin{equation}
    \mathcal{E}\left[\rho\right] = (1-\gamma)\rho+\gamma \, \frac{\mathbb{I}}{2^{N}},
\end{equation}
where $N$ is the number of qubits and $\gamma$, the strength of the noise, is the probability of inducing a completely mixed state. As previously stated, we have evidence that this effect is mostly generated by the $\mathrm{CNOT}$s implemented on the hardware~\cite{Martina2022QMI,IBMQ_ref}. We stress that our model aims to describe all the noise contributions with only one free parameter. We can thus write the quantum dynamical map that evolves the density operator over one cycle of the protocol as
\begin{equation}
    \rho_{n+1} =(1-\gamma) \sum_{k}\Pi_k U\rho_n U^\dagger \Pi_k + \gamma \, \frac{\mathbb{I}}{2^{N}}\,.\label{eq:noisy_evolution_map}
\end{equation}
Thus, since both the quantum map and the depolarizing channel are unital, they commute. This entails that \begin{equation}
    \rho_n = \mathcal{E}[\rho^\mathrm{noiseless}_n] = (1-\gamma)^n \rho^\mathrm{noiseless}_n + \left(1-(1-\gamma)^n\right)\frac{\mathbb{I}}{2^{N}}
\end{equation}
so that
\begin{equation}
    P^{n,\mathrm{noisy}}_{\ket{\phi_k}} = (1-\gamma)^nP^{n,\mathrm{noiseless}}_{\ket{\phi_k}} + \frac{\left(1-(1-\gamma)^n\right)}{2^N}\,.
\end{equation}
As stated previously the quantum dynamical map is unital, thus the identity is a fixed point of the dynamics. This means that the repeated application of the map will bring the system towards the completely mixed state with a timescale 
\begin{equation}
n_\mathrm{noise}\propto \frac{1}{\abs{\ln\left(1-\gamma\right)}}\,.
\end{equation}
For large $n$, the noise hinders partial thermalization and entails the ITT in the whole Hilbert space. Therefore, partial thermalization can be observed only for a time window from the relaxation time of the Markov chain modelling the dynamics of $P^{n}_{\ket{\beta_k}}$ up to the depolarizing time.

\begin{figure}
    \centering
    \includegraphics[width=.49\linewidth]{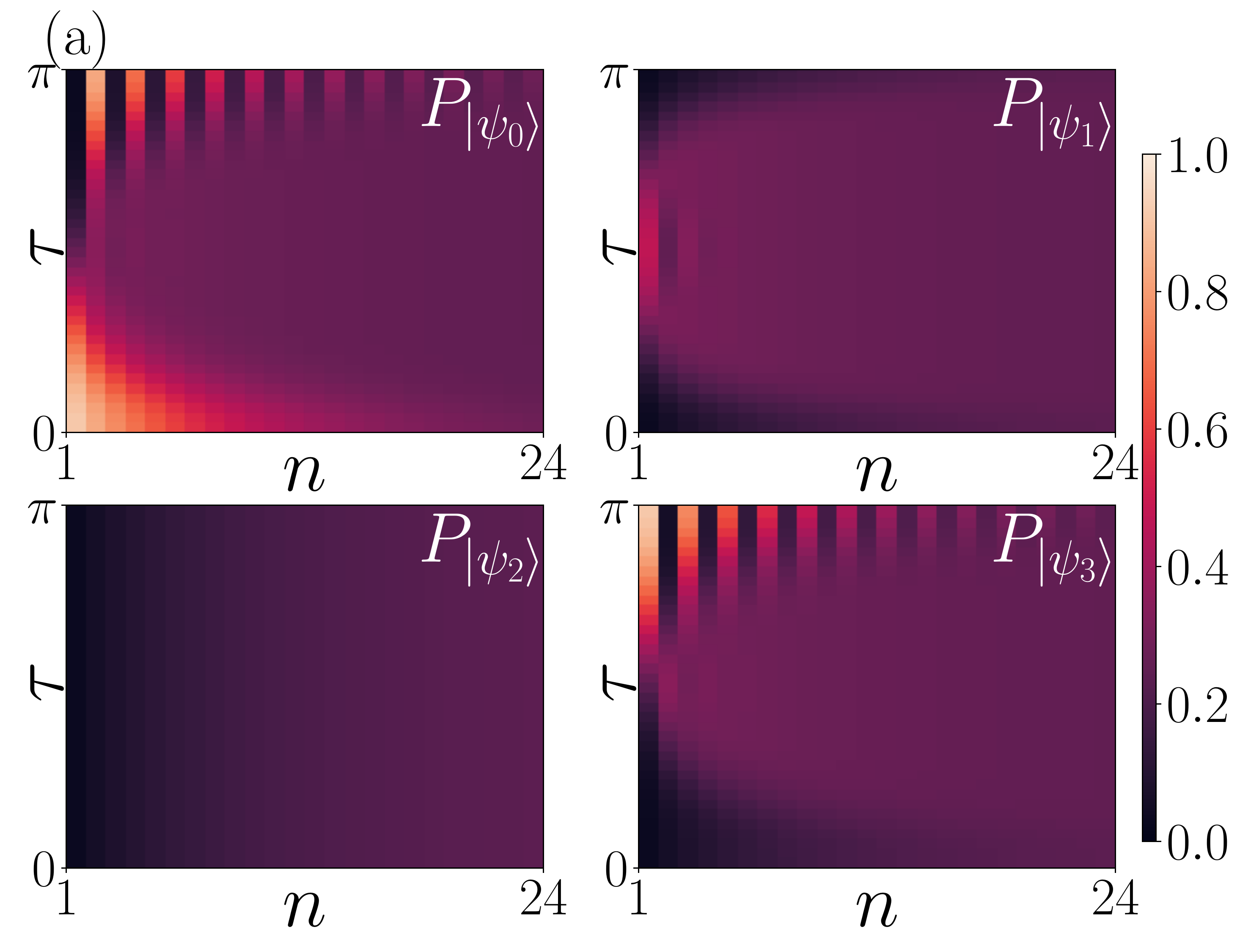}
    \includegraphics[width=.49\linewidth]{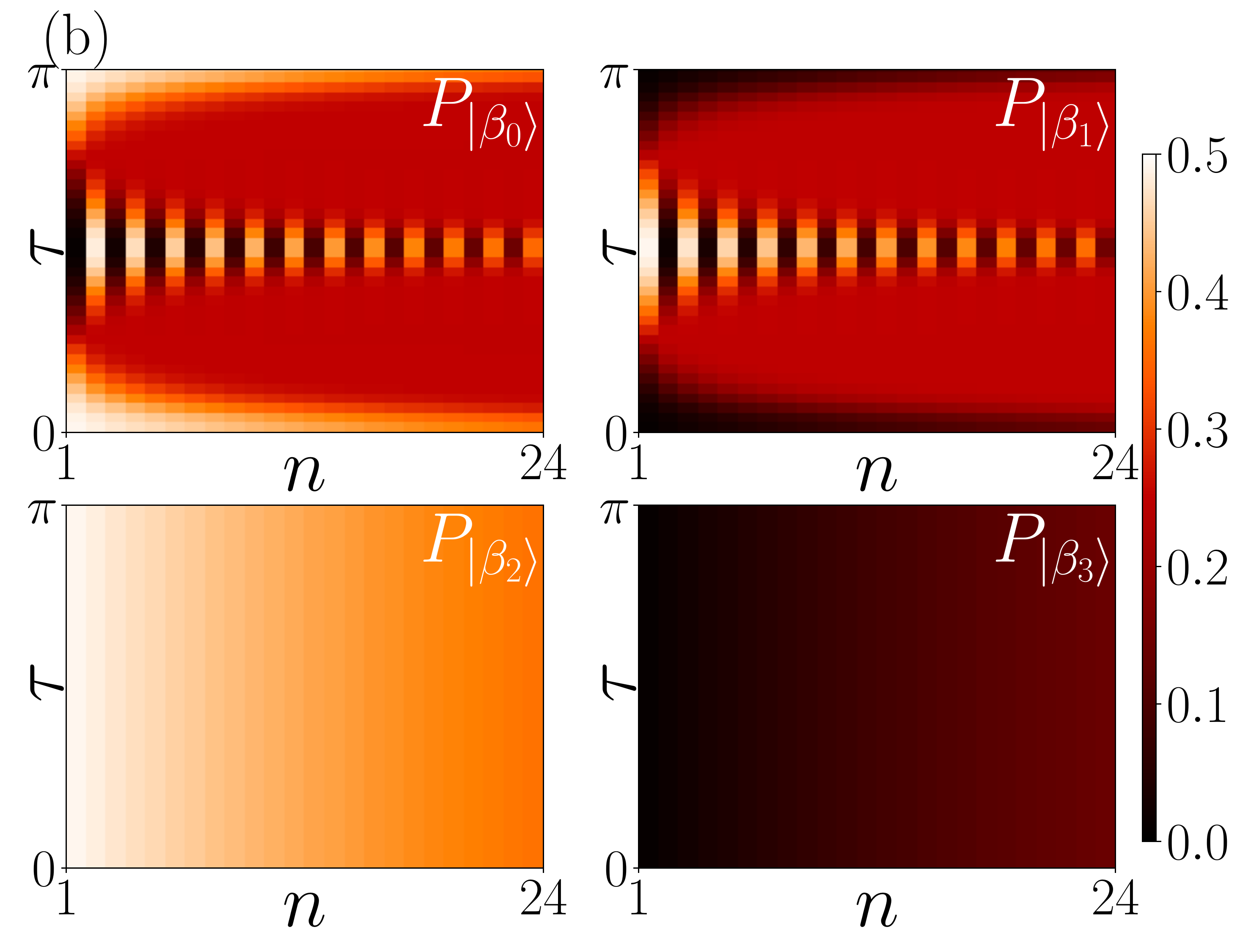}
    \caption{
    Theoretical plots with depolarizing noise. The plotted probabilities have to be compared with the ones in Figs.~\ref{fig:figure2}(a)(b) and~\ref{fig:figure3}(a)(b), with estimated values of the noise strength $\gamma=0.12$ for the panel (a), and $\gamma=0.04$ for panel (b).
    }
    \label{fig:figure6}
\end{figure}

\begin{figure}
    \centering
    \includegraphics[width=.66\linewidth]{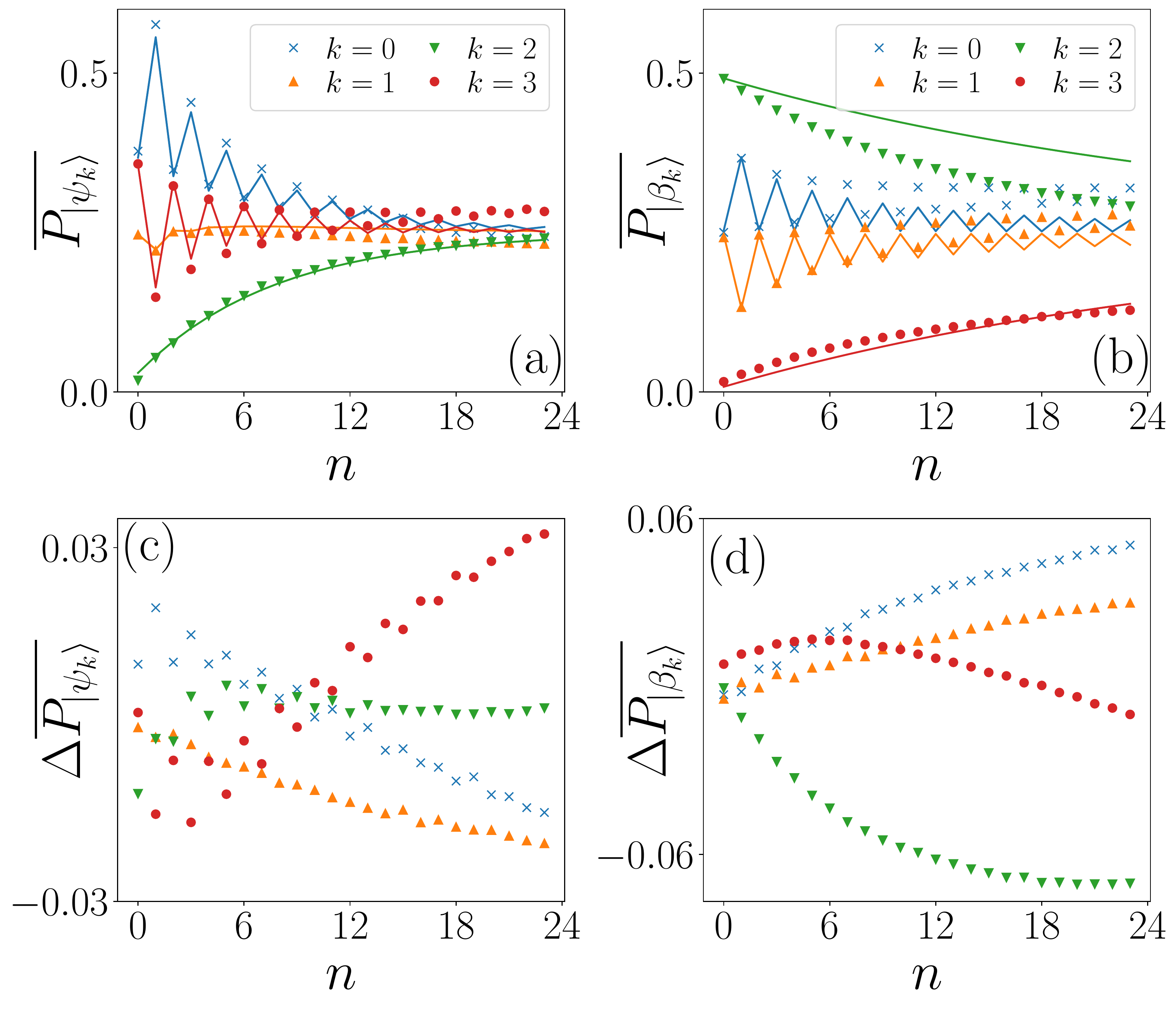}
    \caption{Panels (a),(b): Comparison of the theoretical $\tau$-averaged probabilities $\overline{P}$, solid lines, with data obtained from the quantum hardware, markers, for both the case-studies with a single and two-qubits systems. Panels (c),(d): Absolute difference $\Delta\overline{P}$ of the observed data from the theoretical predictions. Fitted values of $\gamma$: (a),(c) Triplet-singlet measurement basis, $\gamma=0.12$; (b),(d) Bell measurement basis, $\gamma = 0.033$.}
    \label{fig:figure5}
\end{figure}

The value of the parameter $\gamma$ is estimated from the data that are shown in Figs.~\ref{fig:figure2}(b) and~\ref{fig:figure3}(b). For such a purpose, first of all, we compute the $\tau$-average of the measured probabilities, i.e.,
\begin{align}
    \overline{P^n_{\ket{\phi_k}}} = \frac{1}{\pi}\int_0^{\pi} \mathrm{d}\tau \Tr\left[\rho_n\dyad{\phi_k}\right];
    \label{eq: average P}
\end{align}
this allows us to reduce the impact of the errors coming from the local rotations along $x$ that tends to be canceled by averaging. Then, we fit the averaged probabilities~\eqref{eq: average P} with the results of the theoretical simulations for the noisy dynamics that are obtained by evolving $\rho_0$ under the quantum map in Eq.~\eqref{eq:noisy_evolution_map}. The results coming from the numerical simulations of our theoretical model for the noisy protocol are shown in Fig.~\ref{fig:figure6}(a)(b), with the estimated $\gamma$ fitted over the data of Figs.~\ref{fig:figure2}(b) and~\ref{fig:figure3}(b), respectively. A very good agreement is found when comparing the result of noisy simulations in panels Fig.~\ref{fig:figure6}(a)(b) with the data in Figs.~\ref{fig:figure2}(b) and~\ref{fig:figure3}(b), especially for $n\in[1,15]$. This provides us evidence that, despite its simplicity, the noise model introduced in Eq.~\eqref{eq:noisy_evolution_map} is able to capture the main features of the real noise affecting the quantum hardware.

In particular, we have found that in the case-study with the single-triplet measurement basis the strength of the noise $\gamma$ is $\approx 0.12$, meaning that after $n_\mathrm{noise}\approx 8$ the effect of partial thermalization is hidden by the noise. In fact, by inspecting Fig.~\ref{fig:figure5}(a), all the $\tau$-averaged probabilities are equal for $n\approx 24$, which entails that the system is in a completely mixed state. On the other hand, the repeated quantum measurements through a Bell basis is less noisy. The strength of the noise, indeed, is estimated to $\gamma \approx 0.033$, which means that $n_\mathrm{noise} \approx 30$. This is reflected in Fig.~\ref{fig:figure5}(b) where the $\tau$-averaged probabilities for $n=24$ show that the state of the monitored quantum system is not yet completely thermalized to infinite temperature within the whole Hilbert space. Moreover, from our estimation of the single-cycle evolution times, we can infer the decay rate frequency $\Gamma = \gamma/\Delta t$. In particular, we find $\Gamma = 0.02\ \mathrm{MHz}$ for the Bell's measurement basis and $\Gamma = 0.04\ \mathrm{MHz}$ for the monitoring over the singlet-triplet measurement basis. This means that in the singlet-triplet measurement basis the decay rate is two times faster than measuring on the Bell's basis. For the latter case, indeed, we can halve the number of CNOTs to be employed.

As a remark, we stress that only one free-parameter is needed to theoretically reproduce the noisy data. This suggests that recently introduced mitigation techniques, such as the so-called zero noise extrapolation~\cite{Temme2017PRL,Li2017PRX,Kandala2019Nature}, can actually provide a good solution to extract meaningful results from NISQ era intermediate scale quantum devices, with a noise strength that is still far from the fault-tolerant error correction threshold~\cite{AharonovProceedings97} as the one provided by IBMQ. 
Finally, also notice that other possible dissipation models could be applied to explain the noisy data. However, in the cases we are considering, we expect that any model that accounts for extra effects due to noise sources, as for example environmental decoherence and spurious time-dependence of the circuit parameters, would result overall in an exponential decay. Albeit small, such additional effects would contribute to a different relaxation time-scale. In our model, we chose to include only the fastest time-scale, by fitting the single parameter $\gamma$.

\section{Conclusions}

In this paper, we have observed, to our knowledge for the first time, the phenomena of partial and infinite-temperature thermalization on quantum hardware. Specifically, by resorting to IBM quantum machines, \emph{ibm\_lagos} an IBM Quantum Falcon r5.11H processor, we have implemented a sequence of projective measurements, between unitary evolutions, on single-qubit and two-qubit systems. Then, we analyzed the behavior of the monitored system as a function of the number of projections (in our experiments, $n\in[1,32]$ and $n\in[1,24]$ for the single-qubit and two-qubit systems respectively). As predicted in the theoretical paper~\cite{Gherardini2021PRE}, there exist parameter ranges such that partial (infinite-temperature) thermalization occurs depending on the (non)commutativity of $H$, system Hamiltonian, and $\mathcal{O}$, measurement observable. Moreover, we have found that on a quantum hardware the noise is the main obstacle to the long-time stability of partially thermalized states. Yet, we were able to detect them (i.e., partial thermalization) at intermediate time scales that were still resilient to the detrimental effect of the noise.

Our findings are not exclusively aimed at the demonstration of thermalization induced by a sequence of projective measurements on an quantum platform, but they are also expected to have some technological applications for quantum state preparation, provided the system Hamiltonian and the measurement observable. In fact, we have here demonstrated that repeated quantum measurements may make the system converge to unconventional quantum states (neither thermal nor ground states) that are described by a finite effective temperature in each subspace defined by the measurement observable. Hence, these states are not just diagonal in the observable basis but they are block-diagonal, meaning that some coherence terms in such basis are preserved asymptotically in the large $n$ limit.

Finally, as outlooks, it would be worth extending our analysis to the continuous monitoring of quantum many-body systems via a sequence of quantum measurements of local observables, say single spin measurements and to explore the role of symmetries in the Lindblad master equation describing the dynamics \cite{Liang2014PRA}. In this way, connections with numerical results of the recent literature on measurement-induced phase transitions might be determined, and analytical derivations could be accordingly provided.

\section*{Acknowledgements}

We acknowledge the use of IBM Quantum services for this work~\cite{IBMQ_ref}. The views expressed are those of the authors and do not reflect the official policy or position of IBM or the IBM Quantum team. In this paper, we used \textit{ibm\_lagos} which is a IBM Quantum Falcon r5.11H processor. S.G. acknowledges financial support from the MISTI Global Seed Funds MIT-FVG Collaboration Grant ``Non-Equilibrium Thermodynamics of Dissipative Quantum Systems (NETDQS)'', and the PNRR MUR project PE0000023-NQSTI.

\section*{Data availability}

The data that supports the findings of this study are available from the authors upon reasonable request.

\appendix

\section{Hardware properties}
\label{app:Hardware}

In this section we provide some experimental details of the quantum device. In table~\ref{tab:hardware_properties} we report the characterization data of the IBM Quantum processor at the time when the simulations where performed. We refer the reader to the official IBM Quantum site for further details~\cite{IBMQ_ref}. In Sec.~\ref{sec:III} the experiments are implemented on the sixth qubit $q_6$ of the hardware, while in Sec.~\ref{sec:IV} they are executed on the qubits $q_1$ and $q_2$.

\begin{table}
\centering
\begin{tabular}{||c|c||}\hline  Quantum hardware properties & Estimated value \\ \hline \hline
         Duration single qubit gates &  $35$ ns\\ 
         Duration $\mathrm{CNOT}$ & $327$ ns\\ 
         Readout length & $704$ ns\\ \hline
         Relaxation time $T_1$ qubit $q_1$ & $76.4\ \mu s $\\
         Decoherence time $T_2$ qubit $q_1$ & $98.0\ \mu s $\\ \hline
         Relaxation time $T_1$ qubit $q_2$ & $167.1\ \mu s $\\
         Decoherence time $T_2$ qubit $q_2$ & $130.0\ \mu s $\\ \hline
          Relaxation time $T_1$ qubit $q_6$ & $ 110.7 \ \mu s $\\
         Decoherence time $T_2$ qubit $q_6$ & $145.1\ \mu s $\\ \hline
         Average $\sqrt{X}, X, R_z(\theta)$ error & $1.7\times10^{-4}$ \\ 
         $\mathrm{CNOT}$ gate error $(q_1,q_2)$ & $6.8\times 10^{-3}$\\ 
         Maximum readout error & $0.01$\\\hline
\end{tabular}
\caption{Characterization data of \textit{ibmq\_lagos} quantum device at the time our experiments were performed.}
\label{tab:hardware_properties}
\end{table}

\printbibliography

\end{document}